\documentclass[preprint, aps, prd]{revtex4-1}

\def\pd{\partial}
\def\ul{\underline}
\def\mc{\mathcal}

\usepackage[dvips]{graphicx}
\usepackage{amssymb}
\usepackage{amssymb,amsmath}
\usepackage{graphicx}
\usepackage{dsfont}
\usepackage{caption}
\usepackage{subcaption}
\makeatother
\begin{document}

\title{Supersymmetric solutions from $N=6$ gauged supergravity}

\author{Parinya Karndumri} \author{Jakkapat Seeyangnok} \email[REVTeX Support:
]{parinya.ka@hotmail.com and jakkapatjtp@gmail.com} 
\affiliation{String Theory and
Supergravity Group, Department of Physics, Faculty of Science,
Chulalongkorn University, 254 Phayathai Road, Pathumwan, Bangkok
10330, Thailand}

\date{\today}
\begin{abstract}
We study supersymmetric solutions in four-dimensional $N=6$ gauged supergravity with $SO(6)$ gauge group. There is a unique $N=6$ supersymmetric $AdS_4$ vacuum with $SO(6)$ symmetry dual to an $N=6$ SCFT in three dimensions. We find a number of domain walls interpolating between this $AdS_4$ vacuum and singular geometries in the IR with $SO(2)\times SO(4)$, $U(3)$, $SO(3)$ and $SO(2)\times SO(2)\times SO(2)$ symmetries. The $SO(3)$ case admits $N=6$ or $N=2$ solutions depending on whether the pseudoscalars are present or not. On the other hand, all the remaining solutions preserve $N=6$ supersymmetry. These solutions describe RG flows from the $N=6$ SCFT to non-conformal field theories driven by mass deformations. In particular, the $SO(2)\times SO(4)$ solution is in agreement with the previously known mass deformations of the dual $N=6$ SCFT. We also give a supersymmetric Janus solution with $SO(2)\times SO(4)$ symmetry, describing two-dimensional conformal defects in the $N=6$ SCFT with unbroken $N=(4,2)$ supersymmetry. Finally, we find an $N=2$ supersymmetric $AdS_2\times H^2$ solution with $SO(2)\times SO(4)$ symmetry and the corresponding domain wall interpolating between this fixed point and the $AdS_4$ vacuum. The solution describes an $AdS_4$ black hole with a magnetic charge and is dual to a twisted compactification of the $N=6$ SCFT on a hyperbolic space $H^2$. We also give a domain wall interpolating between a locally supersymmetric $AdS_4$ and a curved domain wall with $SO(2)\times SO(2)\times SO(2)$ symmetry dual to an RG flow across dimensions from the $N=6$ SCFT to a supersymmetric quantum mechanics.  
\end{abstract}
\maketitle

\section{Introduction}
Supersymmetric solutions of gauged supergravities in various space-time dimensions play an important role in string/M-theory. In the AdS/CFT correspondence \cite{maldacena,Gubser_AdS_CFT,Witten_AdS_CFT}, these solutions provide holographic descriptions of strongly coupled systems such as (non) conformal field theories, conformal defects, $AdS$-black holes and condensed matter physics. In many cases, solutions of lower-dimensional gauged supergravities can be uplifted to ten or eleven dimensions via consistent truncations resulting in complete AdS/CFT dualities in the context of string/M-theory. 
\\
\indent In this paper, we are interested in supersymmetric solutions of four-dimensional $N=6$ gauged supergravity with $SO(6)$ gauge group, see \cite{Ortin_SUSY_solution} for all time-like supersymmetric solutions in ungauged $N=6$ supergravity. This has been constructed in \cite{Mario_N6} in the embedding tensor formalism obtained from truncating the maximal $N=8$ gauged supergravity \cite{N8_4D_gauged_embedded}, see also \cite{de_Wit_Nicolai,Gauged_maximal_Henning,magnetic_charge_gauged}. The $N=6$ gauged supergravity has been shown to admit a unique $N=6$ supersymmetric $AdS_4$ vacuum, with the full $SO(6)$ symmetry unbroken, dual to an $N=6$ superconformal field theory (SCFT) in three dimensions. The uniqueness of the $N=6$ $AdS_4$ vacua has also been shown in a recent result on supersymmetric $AdS$ vacua \cite{Severin_Maximal_AdS}. It has been pointed out in \cite{Mario_N6} that this $AdS_4$ fixed point describes a truncation of type IIA theory on $CP^3$, so the $AdS_4$ vacuum can be uplifted to $AdS_4\times CP^3$ geometry in type IIA theory. This truncation has been studied long ago in \cite{Pope_IIA_CP3} in which the full mass spectrum has also been given, for more recent studies see for example \cite{AdS4_CP3_1,AdS4_CP3_2,AdS4_CP3_3,AdS4_CP3_4}.
\\
\indent The very first example of the dual $N=6$ SCFT from type IIA theory has been given in \cite{ABJM}. In general, SCFTs in three dimensions take the form of Chern-Simons-Matter (CSM) theories since the usual gauge theories with Yang-Mills gauge kinetic terms are not conformal. A number of these SCFTs with different numbers of supersymmetries have already been constructed, see \cite{Bena,BL1,BL2,BL3,Gustavsson,Basu_Harvey,Schwarz_3D_SCFT,ABJ,N5_6_3D_SCFT,ST,BL4,LP,Origin_3_algebra,
Eric_3D_supercon,Chen,Eric_M2_brane,Eric_Mbrane_SUGRA,MFM,BB,Honda} for an incomplete list. These SCFTs arise as world-volume theories of M2-branes on various transverse spaces and play an important role in understanding the dynamics of M2-branes. Supersymmetric solutions of four-dimensional gauged supergravities are expected to be very useful for their holographic descriptions at least in the large-$N$ limit. 
\\
\indent Various types of supersymmetric solutions from gauged supergravities have been considered and given interpretations in terms of the corresponding dual field theories. We will study these solutions in $N=6$ gauged supergravity beginning with supersymmetric domain walls interpolating between the $N=6$ supersymmetric $AdS_4$ vacuum and singular geometries. The solutions describe holographic RG flows from the dual $N=6$ SCFT in the UV to non-conformal phases in the IR arising from mass deformations of the UV $N=6$ SCFT. Similar solutions have been extensively studied in $N=8$ and $N=2$ gauged supergravities, see for example \cite{Warner_membrane_flow,Warner_M_F_theory_flow,Warner_higher_Dflow,Flow_in_N8_4D,4D_G2_flow,
Warner_M2_flow,Guarino_BPS_DW,Elec_mag_flows,Yi_4D_flow}. Solutions in gauged supergravities with $N=3,4,5$ supersymmetries have also been considered recently in \cite{N3_SU2_SU3,N3_4D_gauging,tri-sasakian-flow,orbifold_flow,4D_N4_flows,N5_flow}. This work could hopefully fill the existing gap by providing a number of supersymmetric solutions in $N=6$ gauged supergravity.  
\\
\indent We will also find Janus solutions in the form of $AdS_3$-sliced domain walls interpolating between asymptotically $AdS_4$ spaces. These are holographically dual to two-dimensional conformal defects within the $N=6$ SCFT and break the superconformal symmetry in the three-dimensional bulk to a smaller one on the two-dimensional surfaces. Solutions of this type in other four-dimensional gauged supergravities have previously been studied in \cite{tri-sasakian-flow,orbifold_flow,N5_flow,warner_Janus,N3_Janus,Minwoo_4DN8_Janus,Kim_Janus}. Finally, we will look for $AdS_2\times \Sigma^2$ geometries with $\Sigma^2$ being a Riemann surface together with solutions interpolating between these backgrounds and the supersymmetric $AdS_4$ vacuum. The solutions describe supersymmetric black holes in asymptotically $AdS_4$ space, and a number of these solutions have already been studied in other gauged supergravities in \cite{N5_flow,BH_M_theory1,BH_M_theory2,Kim_AdS2,AdS4_BH1,AdS4_BH2,AdS4_BH3,AdS4_BH4,AdS4_BH5,
Guarino_AdS2_1,Guarino_AdS2_2,Trisasakian_AdS2}. In the dual field theory, the solutions describe RG flows from the $N=6$ SCFT to superconformal quantum mechanics in the IR which play a prominent role in microscopic computation of black hole entropy in asymptotically $AdS_4$ spaces, see for example \cite{Zaffaroni_BH_entropy,BH_entropy_benini,BH_entropy_Passias}. In this context, the superconformal quantum mechanics, or one-dimensional SCFT, is obtained from the $N=6$ SCFT via twisted compactifications on $\Sigma^2$. 
\\
\indent Four-dimensional $N=6$ gauged supergravity has $SO^*(12)$ global symmetry with the maximal compact subgroup $U(6)\sim SU(6)\times U(1)$. There are thirty scalars encoded in $SO^*(12)/U(6)$ coset manifold. The $SO(6)$ gauging of this supergravity can be obtained from a consistent truncation of the maximal $N=8$ gauged supergravity with $SO(8)$ gauge group. The latter is in turn a consistent truncation of eleven-dimensional supergravity on $S^7$ \cite{N8_4D_S7_1,N8_4D_S7_2,N8_4D_S7_3,N8_4D_S7_4,N8_4D_S7_5,N8_4D_S7_6}. The $N=6$ gauged supergravity with $SO(6)$ gauge group can accordingly be uplifted to eleven dimensions via a series of consistent truncations. On the other hand, the $SO(6)$, $N=6$ gauged supergravity is a consistent truncation of type IIA theory on $CP^3$. Therefore, all the solutions given here have known higher-dimensional origins and can be embedded in ten- or eleven-dimensional supergravities. The scalar potential of the $N=6$ gauged supergravity has been analyzed for a long time in \cite{Warner_Potential}. More recently, this gauged supergravity has been rewritten in a more general setting of the embedding tensor formalism in \cite{Mario_N6} in which the fermion-shift matrices and the scalar potential have been given by truncating the $N=8$ theory. In this paper, we first complete the task by extending the truncation to all terms in the bosonic Lagrangian and fermionic supersymmetry transformations. Both of these are of course a relevant part in the present analysis.   
\\
\indent The paper is organized as follows. In section \ref{N6_SUGRA},
we review four-dimensional $N=6$ gauged supergravity with $SO(6)$ gauge group in the embedding tensor formalism. In section \ref{RG_flow}, we study supersymmetric domain wall solutions describing RG flows in the dual $N=6$ SCFT to non-conformal phases in the IR. We then turn to supersymmetric Janus solutions in section \ref{Janus_N6} and finally look for possible supersymmetric $AdS_2\times \Sigma^2$ solutions together with flow solutions interpolating between the $AdS_4$ vacuum and these geometries in section \ref{AdS4_BH}. Conclusions and comments are given in section \ref{conclusion}.

\section{$N=6$ gauged supergravity with $SO(6)$ gauge group}\label{N6_SUGRA}
We first give a review of $N=6$ gauged supergravity in the embedding tensor formalism as described in \cite{Mario_N6}. We will follow most of the convention in \cite{Mario_N6} but with a mostly plus signature for the space-time metric. The only supermultiplet in $N=6$ supersymmetry is the gravity multiplet with the field content
\begin{equation}
(e^{\hat{\mu}}_\mu,\psi_{\mu A},A^{AB}_\mu,A^0_\mu, \chi_{ABC},\chi_A,\phi_{AB}).
\end{equation}
The component fields correspond to the graviton $e^{\hat{\mu}}_\mu$, six gravitini $\psi_{\mu A}$, sixteen vectors $A^{AB}_\mu=-A^{BA}_\mu$ and $A^0_\mu$, twenty-six spin-$\frac{1}{2}$ fields $\chi_{ABC}=\chi_{[ABC]}$ and $\chi_A$ together with fifteen complex scalars $\phi_{AB}=-\phi_{BA}$. Real and imaginary parts of $\phi_{AB}$ are usually called scalars and pseudoscalars, respectively.
\\
\indent In this work, space-time and tangent space indices are denoted by $\mu,\nu,\ldots =0,1,2,3$ and $\hat{\mu},\hat{\nu},\ldots =0,1,2,3$, respectively. Indices $A,B,\ldots=1,2,\ldots, 6$ correspond to the fundamental representation of $SU(6)$ which is in turn a subgroup of the R-symmetry $U(6)\sim SU(6)\times U(1)$. The $30$ real scalars within $\phi_{AB}$ are coordinates of the scalar manifold $SO^*(12)/U(6)$ and can be described by the coset representative in representation $\mathbf{32}$ of $SO^*(12)$ of the form
\begin{equation}
{\mc{V}_M}^{\ul{M}}=\mc{A}^\dagger e^{Y}
 \end{equation}
 with the Cayley matrix
 \begin{equation}
 \mc{A}=\frac{1}{\sqrt{2}}\left(\begin{array}{cc}
 \mathbb{I}_{16}& i \mathbb{I}_{16}\\
 \mathbb{I}_{16}& -i \mathbb{I}_{16}\\
 \end{array}
 \right)
 \end{equation}
 and
 \begin{equation}
Y=\left(\begin{array}{cccc}
                                           0 & 0_{1\times 15} & 0 & \phi_{CD}  \\
                                           0_{15\times 1} & 0_{15\times 15} & \phi_{AB} & \frac{1}{2}\epsilon_{ABCDEF}\bar{\phi}^{EF}\phi_{CD}  \\
                                            0 & \bar{\phi}^{CD} & 0 & 0_{1\times 15}  \\
                                             \bar{\phi}^{AB} & \frac{1}{2}\epsilon^{ABCDEF}\phi_{EF} & 0_{15\times 1} & 0_{15\times 15}  \\
                                         \end{array}
                                       \right).
 \end{equation}
 We also note that $\bar{\phi}^{AB}=(\phi_{AB})^*$.
 \\
 \indent In subsequent analysis, it is useful to define $16\times 16$ submatrices of ${\mc{V}_M}^{\ul{M}}$ by the following identification
 \begin{equation}
 {\mc{V}_{M}}^{\ul{M}}=\left(\begin{array}{cc}
 {\bar{h}_\Lambda}^{\phantom{\Lambda}\ul{\Lambda}}& h_{\Lambda\ul{\Lambda}}\\
 \bar{f}^{\Lambda\ul{\Lambda}}& {f^\Lambda}_{\ul{\Lambda}}\\
 \end{array}
 \right)
 \end{equation}
 in which $\mathbf{f}$, $\mathbf{h}$, $\bar{\mathbf{f}}$ and $\bar{\mathbf{h}}$ satisfy the relations
\begin{eqnarray}
 & &(\mathbf{f}\mathbf{f}^\dagger)^T=\mathbf{f}\mathbf{f}^\dagger,\qquad (\mathbf{h}\mathbf{h}^\dagger)^T=\mathbf{h}\mathbf{h}^\dagger,\qquad \mathbf{f}\mathbf{h}^\dagger-\bar{\mathbf{f}}\mathbf{h}^T=i\mathbb{I}_{16},\nonumber \\
 & &\mathbf{f}^\dagger\mathbf{h}-\mathbf{h}^\dagger\mathbf{f}=-i\mathbb{I}_{16},\qquad \mathbf{f}^T\mathbf{h}-\mathbf{h}^T\mathbf{f}=0\, .
 \end{eqnarray}
 The inverse of ${\mc{V}_M}^{\ul{M}}$ can accordingly be written in terms of $\mathbf{f}$ and $\mathbf{h}$ as
 \begin{equation}
  {\mc{V}_{\ul{M}}}^{M}=\left(\begin{array}{cc}
-i{f^\Lambda}_{\ul{\Lambda}}& ih_{\Lambda\ul{\Lambda}}\\
 i\bar{f}^{\Lambda\ul{\Lambda}}& -i {\bar{h}_\Lambda}^{\phantom{\Lambda}\ul{\Lambda}}\\
 \end{array}
 \right).
 \end{equation}
The sixteen electric gauge fields $A^{AB}$ and $A^0$ combine into $A^\Lambda=(A^0,A^{AB})$. Together with the magnetic dual $A_{\Lambda}$, the gauge fields transform as $\mathbf{32}$ representation of $SO^*(12)$
\begin{equation}
A^M=(A^\Lambda,A_\Lambda).
 \end{equation}
\indent Gaugings are efficiently described by the embedding tensor formalism in which the corresponding gauge generators are defined as
\begin{equation}
X_M={\theta_M}^nt_n
 \end{equation}
  with $t^n$ being the $SO^*(12)$ generators. ${\theta_M}^n$ is called the embedding tensor in term of which the covariant derivative implementing the minimal coupling of various fields can be written as
  \begin{equation}
  D_\mu=\nabla_\mu -gA^M_\mu X_M\, .
  \end{equation}
$\nabla_\mu$ is the usual space-time covariant derivative including (possibly) the local $U(6)$ composite connection. The parameter $g$ is the gauge coupling constant which can be absorbed in the definition of ${\theta_M}^m$.
\\
\indent In $\mathbf{32}$ representation, with $SO^*(12)$ generators ${(t_n)_M}^N$, the embedding tensor can be described by the generalized structure constants
\begin{equation}
{X_{MN}}^ P={\theta_M}^n{(t_n)_N}^P\, .
 \end{equation}
 \indent To define a proper gauging and preserve the full supersymmetry of the ungauged theory, the embedding tensor needs to satisfy the so-called linear and quadratic constraints given respectively by
 \begin{equation}
 {X_{(MN}}^L\Omega_{P)L}=0\qquad \textrm{and}\qquad {\theta_M}^m{\theta_N}^n{f_{mn}}^p+{X_{MN}}^P{\theta_P}^p=0
  \end{equation}
with ${f_{mn}}^p$ being the $SO^*(12)$ structure constants. The former implies that the embedding tensor ${\theta_M}^m$ is in the representation $\mathbf{351}$ of $SO^*(12)$ while the latter gives rise to
\begin{equation}
[X_M,X_N]=-{X_{MN}}^PX_P\, .
 \end{equation}
The gauge generators then form a closed subalgebra for which ${X_{MN}}^P$ act as the corresponding structure constants.
\\
\indent As usual in gauging a supergravity theory, supersymmetry requires some modifications to the ungauged Lagrangian and supersymmetry transformations. These modifications are of first and second order in the gauge coupling constant and can be written in term of the so-called T-tensor
\begin{equation}
{T_{\ul{M}\ul{N}}}^{\ul{P}}={\mc{V}_{\ul{M}}}^M{\mc{V}_{\ul{N}}}^N{\mc{V}_P}^{\ul{P}}{X_{MN}}^P\, .\label{T_def}
\end{equation}
The bosonic Lagrangian of the $N=6$ gauged supergravity can be written as
\begin{eqnarray}
e^{-1}\mc{L}&=&\frac{1}{2}R-\frac{1}{24}P_{\mu ABCD}P^{\mu ABCD}
-\frac{i}{4}\left(\mc{N}_{\Lambda \Sigma}F^{+\Lambda}_{\mu\nu}F^{+\Sigma \mu\nu}-\overline{\mc{N}}_{\Lambda \Sigma}F^{-\Lambda}_{\mu\nu}F^{-\Sigma \mu\nu}\right)-V\, .\nonumber \\
& &
 \end{eqnarray}
\indent The scalar kinetic term is given in term of the vielbein $P_\mu^{ABCD}=(P_{\mu ABCD})^*$ on the $SO^*(12)/U(6)$ coset which is defined by
\begin{eqnarray}
P_\mu^{ABCD}&=&\mc{V}^{ABM}D_\mu {\mc{V}_M}^{CD}\nonumber \\
&=&i\left(\bar{f}^{\Lambda AB}D_\mu {\bar{h}_\Lambda}^{\phantom{\Lambda}CD}-{\bar{h}_\Lambda}^{\phantom{\Lambda}AB}D_\mu \bar{f}^{\Lambda CD}\right).
\end{eqnarray}
The scalar matrix appearing in the gauge kinetic terms is given by
\begin{equation}
\mc{N}_{\Lambda \Sigma}=-{\bar{h}_\Lambda}^{\phantom{\Lambda}\ul{\Lambda}}(f^{-1})_{\ul{\Lambda}\Sigma}
\end{equation}
with $\overline{\mc{N}}_{\Lambda \Sigma}$ being its complex conjugate. The complex self-dual and anti-self-dual gauge field strengths are defined by
\begin{equation}
F^{\pm\Lambda}_{\mu\nu}=\frac{1}{2}\left(F^\Lambda_{\mu\nu}\pm \frac{i}{2}\epsilon_{\mu\nu\rho\sigma}F^{\Lambda \rho\sigma}\right)
 \end{equation}
with $F^\Lambda_{\mu\nu}$ given by
\begin{equation}
F^\Lambda_{\mu\nu}=\pd_{\mu}A^\Lambda_\nu-\pd_\nu A_\mu^\Lambda+{X_{\Gamma\Sigma}}^\Lambda A^\Gamma_\mu A^\Sigma_\nu\, .
\end{equation}
\\
\indent The scalar potential is obtained from the fermion-shift matrices as follows
\begin{equation}
V=-2S_{AB}S^{AB}+\frac{1}{36}{N_A}^{BCD}{N^A}_{BCD}+\frac{1}{6}{N_A}^B{N^A}_B\, .
\end{equation}
We note that upper and lower $SU(6)$ indices are related by complex conjugation. In terms of the various components of the T-tensor with the spliting of indices $\ul{\Lambda},\ul{\Sigma},\ldots$ as $(0,[AB])$, we have
\begin{eqnarray}
S_{AB}&=&\frac{\sqrt{2}}{5}{T_{C(A,B)E}}^{CE},\qquad N_{AB}=-\frac{8\sqrt{2}}{3}{T_{C[A,B]E}}^{CE},\nonumber \\
{N^A}_B&=&-2\sqrt{2}{T_{0,BC}}^{AC},\qquad {N^A}_{BCD}=-2\sqrt{2}{T_{[CD,B]E}}^{AE}-\frac{1}{4}\delta^A_{[B}N_{CD]}\, .
\end{eqnarray}
\indent The fermionic supersymmetry transformations, with all fermionic fields vanishing, are given by
\begin{eqnarray}
\delta \psi_{\mu A}&=&D_\mu \epsilon_A-S_{AB}\gamma_\mu \epsilon^B-\frac{1}{4\sqrt{2}}\hat{F}^+_{\rho\sigma AB}\gamma^{\rho\sigma}\gamma_\mu \epsilon^B,\\
\delta \chi_A&=&-\frac{1}{4!}\epsilon_{ABCDEF}P^{BCDE}_\mu \gamma^\mu \epsilon^F+{N^B}_A\epsilon_B-\frac{1}{2\sqrt{2}}\hat{F}^+_{\mu\nu}\gamma^{\mu\nu}\epsilon_A,\\
\delta \chi_{ABC}&=&-P_{\mu ABCD}\gamma^\mu \epsilon^D+{N^D}_{ABC}\epsilon_D-\frac{3}{2\sqrt{2}}\hat{F}^+_{\mu\nu [AB}\epsilon_{C]}\, .
\end{eqnarray}
We note here the chiralities of the fermionic fields
\begin{equation}
\gamma_5\psi_{\mu A}=-\psi_{\mu A},\qquad \gamma_5\chi_{ABC}=-\chi_{ABC},\qquad \gamma_5\chi_A=-\chi_A
\end{equation}
with $\psi^A_\mu$, $\chi^{ABC}$ and $\chi^A$ having opposite chiralities. The tensors $\hat{F}^+_{\mu\nu AB}=(\hat{F}_{\mu\nu}^{- AB})^*$ can be obtained from
\begin{equation}
\hat{F}^{-AB}_{\mu\nu}={\mc{V}_M}^{AB}G^{- M}_{\mu\nu}
\end{equation}
with
\begin{equation}
G^M_{\mu\nu}=\left(\begin{array}{c}
 F^\Lambda_{\mu\nu}\\
 G_{\Lambda \mu\nu}\\ 
 \end{array}\right)
 \end{equation}
 and $G_{\Lambda \mu\nu}=i\epsilon_{\mu\nu\rho\sigma}\frac{\pd \mc{L}}{\pd F^\Lambda_{\rho\sigma}}$. Similarly, we have $\hat{F}^+_{\mu\nu}=({\mc{V}_M}^0G^{-M}_{\mu\nu})^*$.
 \\
 \indent The covariant derivative of $\epsilon_A$ is defined by
 \begin{equation}
D_\mu \epsilon_A=\pd_\mu\epsilon_A+\frac{1}{4}{\omega_\mu}^{ab}\gamma_{ab}\epsilon_A+\frac{1}{2}{Q_{\mu A}}^B\epsilon_B\, .
\end{equation}
The connection ${Q_{\mu A}}^B$ is given by
\begin{equation}
{Q_{\mu A}}^B=\frac{2i}{3}\left(h_{\Lambda AC}\pd_\mu \bar{f}^{\Lambda AB}-{f^\Lambda}_{AC}\pd_\mu {\bar{h}_\Lambda}^{\phantom{\Lambda}BC}\right)-gA^M_\mu {Q_{MA}}^B
\end{equation}
with ${Q_{MA}}^B$ obtained from
\begin{equation}
{Q_{M AB}}^{CD}={\mc{V}_{AB}}^P{X_{MP}}^N{\mc{V}_N}^{CD}
\end{equation}
by the relation ${Q_{MAB}}^{CD}=4\delta^{[C}_{[A}{Q_{MB]}}^{D]}$.
\\
\indent In general, both electric and magnetic gauge fields can participate in the gaugings leading to many possibilities of viable gauge groups. However, in this work, we will only consider $SO(6)$ gauge group embedded electrically in $U(6)\subset SO^*(12)$. This gauging only involves electric gauge fields $A^{AB}$. In this case, we have
\begin{equation}
{X_{I_1J_1,I_2J_2}}^{I_3J_3}=4g\delta^{[I_3}_{[I_1}\delta_{I_2][J_2}\delta^{J_3]}_{J_2]}\qquad \textrm{and}\qquad {{X_{I_1J_1}}^{I_3J_3}}_{I_2J_2}=-{X_{I_1J_1,I_2J_2}}^{I_3J_3}
 \end{equation}
with all remaining components vanishing. In particular, there are no ${{X^\Lambda}_M}^N$ components which couple to magnetic gauge fields.
\\
\indent With the splitting of indices $\Lambda,\Sigma,\ldots=(0,[IJ])$, we find from the definition \eqref{T_def} that
\begin{eqnarray}
{T_{AB}}^{CD}&=&-\frac{1}{2}{f^{I_1J}}_0({f^{JJ_1}}_{AB}{\bar{h}_{I_1J_1}}^{\phantom{I_1J_1}CD}+h_{I_1J_1,AB}\bar{f}^{JJ_1,CD}),\\
{T_{EF,AB}}^{CD}&=&-\frac{1}{2}{f^{I_1J}}_{EF}({f^{JJ_1}}_{AB}{\bar{h}_{I_1J_1}}^{\phantom{I_1J_1}CD}+h_{I_1J_1,AB}\bar{f}^{JJ_1,CD}).
 \end{eqnarray}
From these, it is straightforward to obtain all the fermion-shift matrices and the scalar potential.
\\
\indent In subsequent sections, we will look for various types of supersymmetric solutions to this $N=6$ gauged supergravity with $SO(6)$ gauge group. It has been shown in \cite{Mario_N6} that this gauged supergravity admits a supersymmetric $N=6$ $AdS_4$ vacuum with the cosmological constant $V_0=-48g^2$ when all scalars vanish. According to the AdS/CFT correspondence, this is dual to an $N=6$ SCFT in three dimensions. We will find solutions that are asymptotic to this $AdS_4$ geometry and can be interpreted as different types of deformations of the dual $N=6$ SCFT.

\section{Holographic RG flows}\label{RG_flow}
We first consider holographic RG flow solutions in the form of domain walls interpolating between the supersymmetric $AdS_4$ vacuum and another $AdS_4$ vacuum (if exists) or a singular geometry. These solutions correspond respectively to RG flows of the dual UV $N=6$ SCFT to another conformal fixed point or to a non-conformal phase in the IR.
\\
\indent The metric ansatz is taken to be
\begin{equation}
ds^2=e^{2A(r)}dx^2_{1,2}+dr^2
\end{equation}
with $dx^2_{1,2}$ being the flat metric on three-dimensional Minkowski space. Scalar fields are allowed to depend only on the radial coordinate $r$ with all the other fields set to zero.

\subsection{Solutions with $SO(2)\times SO(4)$ symmetry}
We first consider a simple case of solutions with $SO(2)\times SO(4)$ symmetry. The embedding of $SO(6)$ implies that the scalars $\phi_{AB}$ transform as an adjoint representation of $SO(6)$. There is one singlet scalar under $SO(2)\times SO(4)\subset SO(6)$ given explicitly by
\begin{equation}
\phi_{AB}=\phi (\delta_A^1\delta^2_B-\delta^1_B\delta^2_A).
\end{equation}
We will also write
\begin{equation}
\phi=\varphi e^{i\zeta}
\end{equation}
for real scalars $\varphi$ and $\zeta$ depending only on $r$.
\\
\indent By a straightforward computation, we find the tensor $S_{AB}$ of the form
\begin{equation}
S_{AB}=2g\cosh \varphi\delta_{AB}=\frac{1}{2}\mc{W}\delta_{AB}\, .
 \end{equation}
We have introduced the ``superpotential'' $\mc{W}$ for convenience. In general, the function $\mc{W}$ is related to the eigenvalue of $S_{AB}$ corresponding to the unbroken supersymmetry. In the present case, $S_{AB}$ is proportional to the identity matrix indicating that the solutions will preserve either $N=6$ supersymmetry with all $\epsilon_A$ non-vanishing or no supersymmetry at all. Note also that $\mc{W}$ has a critical point at $\varphi=0$ which is the supersymmetric $N=6$ $AdS_4$ vacuum mentioned above.
 \\
 \indent To solve all the BPS conditions, we will, as in other previous works, impose the following projector
\begin{equation}
\gamma_{\hat{r}}\epsilon_A=e^{i\Lambda}\epsilon^A\label{gamma_r_Pro}
 \end{equation}
for a real function $\Lambda$. Throughout the paper, we will use Majorana representation for gamma matrices in which all $\gamma^\mu$ are real, but $\gamma_5$ is purely imaginary. This implies that $\epsilon_A$ and $\epsilon^A$ are related by complex conjugation. Note also that the projector \eqref{gamma_r_Pro} relates the two chiralities of $\epsilon^A$, so the full flow solutions will preserve only half of the original supersymmetry or twelve supercharges.
 \\
 \indent Considering the conditions $\delta\psi_{\mu A}=0$ for $\mu=0,1,2$, we find
 \begin{equation}
 e^{i\Lambda}A'-\mc{W}=0
  \end{equation}
with $'$ denoting $r$-derivatives. This equation gives
 \begin{equation}
 A'= \pm |\mc{W}|\qquad \textrm{and}\qquad e^{i\Lambda}=\pm \frac{\mc{W}}{|\mc{W}|}\, .
 \end{equation}
In what follow, we will write $W=|\mc{W}|$ for convenience. We will also choose the upper signs in order to make the supersymmetric $AdS_4$ critical point correspond to $r\rightarrow \infty$. Since, in this case, the superpotential is real, we simply have
\begin{equation}
A'=4g\cosh\varphi\qquad \textrm{and}\qquad e^{i\Lambda}=1\, .
\end{equation}
The condition $\delta \psi_{r A}=0$ gives the standard Killing spinors of the domain walls
\begin{equation}
 \epsilon_A=e^{\frac{A}{2}}\epsilon_{A(0)}
\end{equation}
for spinors $\epsilon_{A(0)}$ satisfying \eqref{gamma_r_Pro}.
 \\
 \indent Using the projection \eqref{gamma_r_Pro} in the variations $\delta \chi_{ABC}$ and $\delta \chi_A$ gives the following BPS equations
 \begin{equation}
\varphi'=-4g\sinh\varphi\qquad \textrm{and}\qquad  \zeta'=0\, .
  \end{equation}
We have now obtained the BPS equations that solve all the supersymmetry conditions. It can also be readily verified that these equations imply the second-order field equations.
\\
\indent We can analytically solve the above BPS equations with the following solution
\begin{eqnarray}
4gr&=&\ln(1+e^\varphi)-\ln (1-e^\varphi),\\
A&=&\varphi-\ln (1-e^{2\varphi}).
 \end{eqnarray}
We have neglected the integration constants in these equations since they can be removed by shifting the radial coordinate and scaling the $x^{0,1,2}$ coordinates, respectively. As $r\rightarrow \infty$, we find that
\begin{equation}
\varphi\sim e^{-4gr} \sim e^{-\frac{r}{L}}\qquad \textrm{and}\qquad A\sim 4gr\sim \frac{r}{L}
\end{equation}
with $L$ being the $AdS_4$ radius related to the cosmological constant by
\begin{equation}
L=\sqrt{-\frac{3}{V_0}}=\frac{1}{4g}\, .
\end{equation}
We have also taken $g>0$ for convenience.
\\
\indent The behavior of $\varphi$ implies that $\varphi$ is dual to a relevant operator of dimensions $\Delta=1,2$ in the dual SCFT. In addition, the solution is singular at $r\rightarrow 0$ with
\begin{equation}
\varphi\sim \pm \ln (4gr)\qquad \textrm{and}\qquad A\sim \ln(4gr)\, .
\end{equation}
We then find that $\varphi\rightarrow\pm \infty$ near the singularity. From the explicit form of the scalar potential, we have
\begin{equation}
 V\sim -8g^2e^{\pm 2\varphi}\rightarrow -\infty\, .
\end{equation}
By the criterion given in \cite{Gubser_singularity}, we conclude that the singularity is physical. Therefore, the solution decribes an RG flow from the UV $N=6$ SCFT to a non-conformal phase in the IR. The flow is driven by an operator of dimensions $\Delta=1,2$ corresponding to scalar or fermion mass terms in three dimensions. The flow breaks superconformal symmetry but preserves the full $N=6$ Poincare supersymmetry. Moreover, the R-symmetry $SO(6)$ is broken to $SO(2)\times SO(4)$ subgroup. This is precisely in agreement with the field theory result given in \cite{N5_6_3D_SCFT}. We then expect the solution to describe mass deformations of the three-dimensional $N=6$ SCFT.

\subsection{Solutions with $U(3)$ symmetry}
We now consider another residual symmetry namely $U(3)\sim SU(3)\times U(1)\subset SO(6)$. The $U(3)$ generators in the fundamental representation of $SO(6)$ can be written as
\begin{equation}
X=\left(\begin{array}{cc}
A_{3\times 3}& S_{3\times 3}\\
 -S_{3\times 3}& A_{3\times 3}\\
 \end{array}
 \right)\label{U3_gen}
\end{equation}
in which $A_{3\times 3}$ and $S_{3\times 3}$ are anti-symmetric and symmetric $3\times 3$ matrices, respectively. The matrices $A_{3\times 3}$ generate an $SO(3)\subset SU(3)$ which is a diagonal subgroup of $SO(3)\times SO(3)\subset SO(6)$. The $U(1)$ factor corresponds to $S_{3\times 3}=\mathbb{I}_3$. There is only one $U(3)$ singlet scalar given by
\begin{equation}
\phi_{AB}=\left(\begin{array}{cc}
0_{3\times 3}& \phi \mathbb{I}_{3}\\
 -\phi \mathbb{I}_{3}& 0_{3\times 3}\\
 \end{array}
 \right)=\phi J_{AB}\, .\label{U3_singlet}
\end{equation}
The matrix $J_{AB}$ is identified with the Kahler form of $CP^3$ on which the ten-dimensional type IIA theory compactifies \cite{Mario_N6}.
\\
\indent By writting $\phi=\varphi e^{i\zeta}$ and repeating the same analysis as in the previous case, we find the scalar potential
\begin{equation}
V=-24g^2e^{-2\varphi}(1+e^{4\varphi})
\end{equation}
which is exactly the same as that given in \cite{Mario_N6}. As in the $SO(2)\times SO(4)$ case, this potential admits an $N=6$ $AdS_4$ critical point at $\varphi=0$ dual to an $N=6$ SCFT in three dimensions.
\\
\indent The matrix $S_{AB}$ is proportional to the identity
\begin{equation}
S_{AB}=\frac{1}{2}\mc{W}\delta_{AB}
\end{equation}
with a complex superpotential
\begin{equation}
\mc{W}=\frac{1}{2}e^{-3\varphi-i\zeta}\left[(e^{6\varphi}+3e^{2\varphi})(1+e^{i\zeta})+(1+e^{4\varphi})(e^{i\zeta}-1)\right].
\end{equation}
The variations $\delta \chi_A$ and $\delta \chi_{ABC}$ lead to
\begin{equation}
e^{-i\Lambda}(2\varphi'\pm i\sinh (2\varphi) \zeta')=-ge^{-3\varphi}(e^{4\varphi}-1)[1-e^{i\zeta}+e^{2\varphi}(1+e^{i\zeta})]
\end{equation}
which implies $\zeta'=0$. It turns out that $\zeta'=0$ is also required by the field equations. For constant $\zeta=\zeta_0$, we have verified that all the resulting BPS equations are compactible with the field equations. In the following analysis, we will set $\zeta_0=0$ and end up with the BPS equations
\begin{equation}
\varphi'=-e^{-\varphi}(e^{4\varphi}-1)\qquad \textrm{and}\qquad A'=ge^{-\varphi}(3+e^{4\varphi}).
\end{equation}
The solution can be readily obtained
\begin{eqnarray}
A&=&3\varphi-\ln(1-e^{4\varphi}),\\
4gr&=&2\tan^{-1}e^\varphi-\ln(1-e^\varphi)+\ln (1+e^{\varphi}).
\end{eqnarray}
\indent As in the previous case, the solution is asymptotic to the supersymmetric $AdS_4$ with $\varphi$ dual to an operator of dimensions $\Delta=1,2$ while at $r=0$, the solution is singular with
\begin{equation}
\varphi\sim \ln(gr)\qquad \textrm{and}\qquad A\sim 3\varphi\sim 3\ln(gr)
\end{equation}
and
\begin{equation}
\varphi\sim -\ln(gr)\qquad \textrm{and}\qquad A\sim -\varphi\sim \ln(gr).
\end{equation}
Both of these give
\begin{equation}
V\sim -24g^2e^{\pm 2\varphi}\rightarrow -\infty,
\end{equation}
so the two singularities are physical. We can accordingly interpret the solution as a holographic dual of RG flows from the $N=6$ SCFT to non-conformal phases in the IR. The flow preserves $N=6$ Poincare supersymmetry in three-dimensions as in the $SO(2)\times SO(4)$ case but breaks the $SO(6)$ R-symmetry to $U(3)$. It would be interesting to identify the corresponding mass deformations in the dual $N=6$ SCFT similar to the $SO(2)\times SO(4)$ case.

\subsection{Solutions with $SO(2)\times SO(2)\times SO(2)$ symmetry}
To obtain more interesting and more complicated solutions, we consider solutions with a smaller symmetry namely $SO(2)\times SO(2)\times SO(2)\subset SO(6)$ symmetry. There are three complex scalars which are singlets under this $SO(2)\times SO(2)\times SO(2)$. The explicit parametrization of these singlets can be written as
\begin{equation}
\phi_{AB}=\left(\begin{array}{ccc}
\phi_1 i\sigma_2& 0_{2\times 2} & 0_{2\times 2}\\
 0_{2\times 2}& \phi_2 i\sigma_2& 0_{2\times 2} \\
 0_{2\times 2}&0_{2\times 2} & \phi_3 i\sigma_2\\
 \end{array}
 \right) \,
 .\label{SO2_3_singlet}
\end{equation}
By setting
\begin{equation}
\phi_\alpha=\varphi_\alpha e^{i\zeta_\alpha},\qquad \alpha=1,2,3,
\end{equation}
we find the scalar potential
\begin{equation}
V=-16g^2[\cosh(2\varphi_1)+\cosh(2\varphi_2)+\cosh(2\varphi_3)].\label{V_SO2_3}
\end{equation}
It is clearly seen that this potential admits only one critical point at $\varphi_1=\varphi_2=\varphi_3=0$ which is the aforementioned $N=6$ $AdS_4$ vacuum.
\\
\indent The matrix $S_{AB}$ is given by
\begin{equation}
S_{AB}=\frac{1}{2}\left(\begin{array}{ccc}
\mc{W}_1 \mathbb{I}_2& 0_{2\times 2} & 0_{2\times 2}\\
 0_{2\times 2}& \mc{W}_2 \mathbb{I}_2& 0_{2\times 2} \\
 0_{2\times 2}&0_{2\times 2} & \mc{W}_3 \mathbb{I}_2\\
 \end{array}
 \right)
\end{equation}
with
\begin{eqnarray}
\mc{W}_1&=&\frac{1}{2}ge^{-\varphi_1-\varphi_2-\varphi_3}\left[e^{i(\zeta_1-\zeta_2-\zeta_3)}(e^{2\varphi_1}-1)(e^{2\varphi_2}-1)(e^{2\varphi_3}-1)\right.\nonumber \\
& &\left.-(1+e^{2\varphi_1})(1+e^{2\varphi_2})(1+e^{2\varphi_3})\right].
\end{eqnarray}
$\mc{W}_2$ and $\mc{W}_3$ take a similar form with the phase $e^{i(\zeta_1-\zeta_2-\zeta_3)}$ replaced by $e^{i(\zeta_2-\zeta_1-\zeta_3)}$ and $e^{i(\zeta_3-\zeta_1-\zeta_2)}$, respectively.
\\
\indent It turns out that none of these $\mc{W}_{\alpha}$ gives rise to the superpotential in term of which the scalar potential \eqref{V_SO2_3} can be written unless $\zeta_1=\zeta_2=\zeta_3=0$. This is also implied by the consistency between the resulting BPS equations and the field equations. We now set $\zeta_1=\zeta_2=\zeta_3=0$, resulting in $e^{i\Lambda}=\pm 1$, and obtain the following BPS equations
\begin{eqnarray}
\varphi_1'&=&-ge^{-\varphi_1-\varphi_2-\varphi_3}[e^{2(\varphi_1+\varphi_2)}+e^{2(\varphi_1+\varphi_3)}-e^{2(\varphi_2+\varphi_3)}-1], \label{eq1_SO2_3}\\
\varphi_2'&=&-ge^{-\varphi_1-\varphi_2-\varphi_3}[e^{2(\varphi_1+\varphi_2)}+e^{2(\varphi_2+\varphi_3)}-e^{2(\varphi_1+\varphi_3)}-1],\\
\varphi_3'&=&-ge^{-\varphi_1-\varphi_2-\varphi_3}[e^{2(\varphi_1+\varphi_3)}+e^{2(\varphi_2+\varphi_3)}-e^{2(\varphi_1+\varphi_2)}-1],\\
A'&=&ge^{-\varphi_1-\varphi_2-\varphi_3}[e^{2(\varphi_1+\varphi_2)}+e^{2(\varphi_1+\varphi_3)}+e^{2(\varphi_3+\varphi_3)}+1].
\end{eqnarray}
\indent To find the solution to these equations, we first take a linear combination
\begin{equation}
\varphi_1'+\varphi_2'=-2ge^{-\varphi_1-\varphi_2-\varphi_3}(e^{2(\varphi_1+\varphi_2)}-1).
\end{equation}
After changing to a new radial coordinate $\rho$ defined by
\begin{equation}
\frac{d\rho}{dr}=e^{-\varphi_1-\varphi_2-\varphi_3},
\end{equation}
we find
\begin{equation}
\varphi_2=2g\rho-\varphi_1-\frac{1}{2}\ln(e^{4g\rho}+C_2)
\end{equation}
for a constant $C_2$.
\\
\indent Similarly, taking a linear combination $\varphi_1'+\varphi_3'$ gives rise to
\begin{equation}
\varphi_3=2g\rho-\varphi_1-\frac{1}{2}\ln (e^{4g\rho}+C_3).
\end{equation}
Using these results in equation \eqref{eq1_SO2_3}, we find
\begin{equation}
\varphi_1=\frac{1}{4}\ln\left[\frac{e^{4g\rho}(e^{4g\rho}+C_1)}{(e^{4g\rho}+C_2)(e^{4g\rho}+C_3)}\right].
\end{equation}
Finally, with all these results, the solution for $A$ is given by
\begin{equation}
A=g\rho+\frac{1}{4}\ln(e^{4g\rho}+C_1)+\frac{1}{4}\ln(e^{4g\rho}+C_2)+\frac{1}{4}\ln(e^{4g\rho}+C_3).
\end{equation}
\indent We now look at the behavior of the solution as $\varphi_\alpha\sim 0$ which gives $\rho\sim r$ and
\begin{equation}
\varphi_1\sim \frac{1}{4}(C_1-C_2-C_3)e^{-4g\rho},\qquad \varphi_{2,3}\sim -\frac{1}{4}(C_1-C_{3,2})e^{-4g\rho},\qquad A\sim 4g\rho\, .
\end{equation}
This is the expected behavior of the solution asymptotic to the supersymmetric $AdS_4$ vacuum. As in the previous cases, the solution is singular as $4g\rho\rightarrow \ln (-C_\alpha)$. For $C_1\neq C_2\neq C_3$, there are three possibilities:
\begin{itemize}
\item $C_1<C_{2,3}$: In this case, the solution is singular when $4g\rho\rightarrow \ln (-C_1)$ with
\begin{eqnarray}
& &\varphi_1\sim \frac{1}{4}\ln(4g\rho-\tilde{C}_1),\qquad \tilde{C}_1=\ln(-C_1),\nonumber \\
& &\varphi_{2,3}\sim -\varphi_1,\qquad A\sim \varphi_1\, .
\end{eqnarray}
\item $C_2<C_{1,3}$ or $C_3<C_{1,2}$: In this case, we find
\begin{eqnarray}
& &\varphi_1\sim -\frac{1}{4}\ln(4g\rho-\tilde{C}_{2,3}),\qquad \tilde{C}_{2,3}=\ln(-C_{2,3}),\nonumber \\
& &\varphi_{2,3}\sim \varphi_1,\qquad A\sim -\varphi_1\, .
\end{eqnarray}
\end{itemize}
In the first case, we have $\varphi_1\rightarrow -\infty$ and $\varphi_{2,3}\rightarrow \infty$ while in the second case, the solution gives $\varphi_{1,2,3}\rightarrow \infty$. It can be easily verified that all of these behaviors lead to $V\rightarrow-\infty$, so the singularities are physically acceptable. The solution then describes different types of mass deformations within the dual $N=6$ SCFT to non-conformal phases with $SO(2)\times SO(2)\times SO(2)$ symmetry. The solution also preserves $N=6$ Poincare supersymmetry as in the previous cases.

\subsection{Solutions with $SO(3)$ symmetry}
We further reduce the residual symmetry to $SO(3)\subset SO(3)\times SO(3)\subset SO(6)$ generated by the antisymmetric matrices $A_{3\times 3}$ in the upper-left block of \eqref{U3_gen}. There are three singlet scalars parametrized by
\begin{equation}
\phi_{AB}=\left(\begin{array}{cc}
0_{3\times 3}& 0_{3\times 3}\\
0_{3\times 3}& \hat{A}_{3\times 3}\\
\end{array}
 \right)
\end{equation}
with
\begin{equation}
\hat{A}=\left(\begin{array}{ccc}
0& \tilde{\phi}_1& \tilde{\phi}_2\\
-\tilde{\phi}_1&0 & \tilde{\phi}_3\\
-\tilde{\phi}_2& -\tilde{\phi}_3&0 \\
\end{array}
  \right).
\end{equation}
By writting $\tilde{\phi}_\alpha=\varphi_\alpha e^{i\zeta_\alpha}$ with
\begin{equation}
\varphi_1=\Phi\cos\theta,\qquad \varphi_2=\Phi\sin\theta\cos\vartheta,\qquad \varphi_3=\Phi\sin\theta\sin\vartheta
\end{equation}
and
\begin{equation}
\zeta_1=\zeta,\qquad \zeta_2=\zeta+\eta,\qquad \zeta_3=\zeta+\xi,
\end{equation}
we find the scalar potential
\begin{eqnarray}
V&=&-g^2\left[16\cos^4\theta(2+\cosh2\Phi)+16\cosh^4\theta\sin^4\theta(2+\cosh2\Phi) \right.\nonumber \\
& &+16\sin^4\theta\sin^4\vartheta(2+\cosh2\Phi)-\cos^2\theta\sin^2\theta\cos^2\vartheta\times \nonumber\\
& &(\cosh4\Phi-8\cos2\eta\sinh^4\Phi-36\cosh2\Phi-61)+\sin^2\theta\sin^2\vartheta \times \nonumber \\
& &\left[8\sinh^4\Phi(\cos^2\theta\cos2\xi+\cos^2\vartheta\sin^2\theta\cos[2(\eta-\xi)]) \right.\nonumber \\
& &\left.\left.+(61+36\cosh2\Phi-\cosh4\Phi)(\cos^2\theta+\cos^2\vartheta\sin^2\theta) \right]\right].
\end{eqnarray}
\indent In this case, the scalar potential depends on phases of the complex scalars, and the analysis is more complicated. To make the analysis more traceable, we will further truncate to two scalars by setting $\vartheta=0$ and $\xi=-\zeta$. This is equivalent to setting $\tilde{\phi}_3=0$. We now begin with the eigenvalues of the matrix $S_{AB}$. These are of the form, after diagonalization,
\begin{equation}
S^{\textrm{diag}}_{AB}=\textrm{diag}(-2g\cosh\Phi_{\times 4},\frac{1}{2}\mc{W}_+,\frac{1}{2}\mc{W}_-)
\end{equation}
with $\mc{W}_\pm$ given by
\begin{eqnarray}
\mc{W}_\pm&=&2g(\cos 2\eta+2\sin\eta)\sinh^4\frac{\Phi}{2}(\cos4\theta\sin\eta\pm i\sin2\theta)\nonumber \\
& &-\frac{1}{4}g(3+12\cosh\Phi+\cosh2\Phi).
\end{eqnarray}
The corresponding eigenvectors are
\begin{equation}
\hat{\epsilon}_\pm=-\frac{1}{2}\sec2\theta\left(2\cos\eta\sin2\theta\mp \sqrt{3+\cos2\eta+2\cos4\theta\sin^2\eta}\right)\epsilon_5+\epsilon_6.
\end{equation}
\indent The scalar kinetic term is given by
\begin{eqnarray}
\mc{L}_{\textrm{kin}}&=&-\frac{1}{2}G_{\alpha\beta}{\phi^\alpha}'{\phi^\beta}'\nonumber \\
&=&-\Phi'^2-\sinh^2\Phi \theta'^2-\frac{1}{4}\sinh^22\Phi \zeta'^2-\frac{1}{2}\sin^2\theta\sinh^22\Phi\zeta'\eta'\nonumber \\
& &-\frac{1}{4}\sin^2\theta\sinh^2\Phi(3+\cosh2\Phi-2\cos2\theta\sinh^2\Phi)\eta'^2
\end{eqnarray}
with $\phi^\alpha=(\Phi, \theta,\zeta,\eta)$. It is useful to give an explicit form of the inverse of $G_{\alpha\beta}$ here
\begin{equation}
G^{\alpha\beta}=-\frac{1}{2}\left(\begin{array}{cccc}
1 & 0 & 0 & 0 \\
0 & \textrm{csch}^2\Phi & 0 & 0\\
0 & 0 & -\textrm{sech}^2\Phi+\textrm{csch}^2\Phi\sec^2\theta & -\textrm{csch}^2\Phi \sec^2\theta \\
0 & 0 & -\textrm{csch}^2\Phi \sec^2\theta & 4\textrm{csc}^22\theta\textrm{csch} \Phi
\end{array}\right).
\end{equation}
The scalar potential can be written in term of the real superpotential $W=|\mc{W}_+|=|\mc{W}_-|$ as
\begin{eqnarray}
V&=&-2G^{\alpha\beta}\frac{\pd W}{\pd \phi^\alpha}\frac{\pd W}{\pd \phi^\beta}-3 W^2 \nonumber \\
&=&g^2\left[\cos^2\theta\sin^2\theta\left(\cosh4\Phi-8\cos2\eta\sinh^4\Phi-36\cosh2\Phi-61\right) \right.\nonumber \\
& &\left.-4(3+\cos4\theta)(2+\cosh2\Phi)\right].
\end{eqnarray}
\indent After setting $\epsilon^{1,2,3,4}=0$ and imposing the projection conditions
\begin{equation}
\gamma_{\hat{r}}\epsilon_{\pm}=e^{\pm i\Lambda}\epsilon^{\pm}\qquad \textrm{with}\qquad e^{\pm i\Lambda}=\frac{\mc{W}_\pm}{W},
\end{equation}
we find the following BPS equations
 \begin{eqnarray}
 \Phi'&=&\frac{1}{16W}g^2\left[8\sinh^3\Phi\cosh\Phi(\cos2\eta+2\cos4\theta\sin^3\eta) 
-30\sin2\Phi-\sinh4\Phi \right],\nonumber \\
& &\\
\theta'&=&-\frac{1}{W}g^2\sin^2\eta\sin4\theta\sinh^2\Phi,\\
\zeta'&=&\frac{2}{W}g^2\sin2\eta\sin^2\theta\sinh^2\Phi,\\
\eta'&=&-\frac{2}{W}g^2\sin2\eta\sinh^2\Phi,\\
A'&=&W\, .
 \end{eqnarray}
The flow equations for the scalars can be written in a compact form as
\begin{equation}
{\phi^\alpha}'=2G^{\alpha\beta}\frac{\pd W}{\pd \phi^\beta}\, .
 \end{equation}
It can straightforwardly be verified that all these equations satisfy the second-order field equations. We also see that from these equations, there is only one supersymmetric critical point, with $\Phi'=\theta'=\zeta'=\eta'=0$, at $\phi^\alpha=0$.
\\
\indent We note that although the superpotential and the scalar potential do not depend on $\zeta$, $\zeta'$ is still non-vanishing due to the mixed terms between $\zeta$ and $\eta$ in the matrix $G_{\alpha\beta}$. Moreover, further truncations such as $\eta=0$ or $\theta=0$ will lead to the BPS equations in the case of $N=6$ supersymmetry with all the six eigenvalues of $S_{AB}$ leading to
\begin{equation}
\mc{W}=4g\cosh\Phi\, .
\end{equation}
This is very similar to the $N=5$ gauged supergravity studied in \cite{N5_flow} in which the differences in the phases of the scalars are crucial for breaking the original supersymmetry to lower amount.
\\
\indent We now look at the solution to the above equations. Combining $\eta'$ and $\theta'$ equations gives
\begin{equation}
\frac{d\theta}{d\eta}=\frac{1}{4}\sin4\theta\tan\eta
\end{equation}
with the solution given by
\begin{equation}
\cot2\theta=C_1\cos\eta\, .
\end{equation}
 Similarly, taking the combination between $\zeta'$ and $\eta'$ equations leads to
\begin{equation}
\frac{d\zeta}{d\eta}=-\sin^2\theta\, .
\end{equation}
After using the above solution for $\theta$, we find the solution
\begin{equation}
\zeta=\zeta_0-\frac{\eta}{2}+\frac{\sqrt{C_1^2+\sec^2\eta}\cos\eta \tan^{1}\frac{\sqrt{2}C_1\sin\eta}{\sqrt{2+C_1^2(1+\cos2\eta)}}}{\sqrt{4+2C_1^2(1+\cos2\eta)}}
\end{equation}
for constant $\zeta_0$.
\\
\indent Combining $\Phi'$ and $\eta'$ equations by taking into account all the previous results together with a redefinition
\begin{equation}
\tilde{\Phi}=\sinh\Phi,
\end{equation}
we find
\begin{equation}
\frac{d\tilde{\Phi}}{d\eta}=\textrm{csc}2\eta(1+\tilde{\Phi}^2)\left(\frac{\tilde{\Phi}\tan^2\eta}{C_1^2+\sec^2\eta}+\frac{2}{\tilde{\Phi}}\right)
\end{equation}
whose solution is given by
\begin{eqnarray}
\frac{\tilde{\Phi}^2}{4}=-\frac{1+C_1^2\cos^2\eta-C_2\sqrt{(1+\cos2\eta)(2+C_1^2(1+\cos2\eta))}}{3+4C_1^2\cos^2\eta+\cos2\eta-4C_2\sqrt{(1+\cos2\eta)(2+C_1^2(1+\cos2\eta))}}\, .
\end{eqnarray}
\indent Using all these results in $\eta'$ equation, we find the solution for $\eta(r)$ implicitly from
\begin{eqnarray}
8gr&=&\sinh^{-1}\left[2C_2\sqrt{\frac{\Xi-1}{(1+C_1^2)[2+(C_1^2-4C_2^2)(1+\Xi)]}}\right]\nonumber \\
& &-\tanh^{-1}\sqrt{\frac{(1+C_1^2-4C_2^2)(1+\Xi)}{\Xi-1}}
\end{eqnarray}
in which we have defined
\begin{equation}
\Xi=\cos2\eta\, .
\end{equation}
\indent Finally, we can find the solution for $A(\Xi)$ as
\begin{eqnarray}
A&=&\frac{1}{4}(\tanh^{-1}\alpha_+-\tanh^{-1}\alpha_-)-\frac{1}{2}\tanh^{-1}\left[2C_2\sqrt{\frac{\Xi+1}{2+C_1^2(1+\Xi)}}\right]\nonumber \\
 & &-\frac{1}{8}\ln\left[4C_1^4(1+\Xi)^2+(3+\Xi)^2-4(1+\Xi)[8C_2^2+C_1^2(4C_2^2(1+\Xi)-3-\Xi)]\right]\nonumber \\
 & &+\frac{1}{4}\ln[2+(1+\Xi)(C_1^2-4C_2^2)]
\end{eqnarray}
with $\alpha_\pm$ defined by
\begin{equation}
\alpha_\pm=\sqrt{-\frac{2+C_1^2(1+\Xi)}{(1+\Xi)\left[1+C_1^2-4C_2(2C_2+\pm \sqrt{4C_2^2-C_1^2-1})\right]}}\, .
\end{equation}
\indent The solution preserves $N=2$ supersymmetry and breaks the $SO(6)$ R-symmetry to $SO(3)$. The solution is singular when  
\begin{equation}
\cos^2\eta=-\frac{1}{1+2C_1^2-8C_2^2\pm 4C_2\sqrt{4C_2^2-C_1^2-1}}
\end{equation}
This gives $\tilde{\Phi}\rightarrow \pm \infty$ or $\Phi\rightarrow \pm \infty$ which in turn leads to 
 \begin{equation}
 V\rightarrow g^2e^{4|\Phi|}\cos^2\theta \sin^2\eta\sin^2\theta\, .
 \end{equation}
We can see that the scalar potential is unbounded from above, $V\rightarrow +\infty$, unless $\theta=0$ or $\eta=0$ both of which give the $N=6$ solution as previously mentioned. Therefore, the IR singularities of the $N=2$ solutions are unphysical by the criterion of \cite{Gubser_singularity}.
\section{Supersymmetric Janus solutions}\label{Janus_N6}
In this section, we consider supersymmetric Janus solutions in the form of curved domain walls. The solutions can be obtained from an $AdS_3$-sliced domain wall ansatz
\begin{equation}
ds^2=e^{2A}(e^{\frac{2\xi}{\ell}}dx^2_{1,1}+d\xi^2)+dr^2\, .
\end{equation}
Since the analysis closely follows that given in \cite{warner_Janus}, see also \cite{N3_Janus}, we will not repeat all the detail here but mainly review relevant results for deriving the corresponding BPS equations.
\\
\indent Compared to the RG flow case, the BPS equations will be modified by the curvature of the three-dimensional slices. In addition, as pointed out in \cite{warner_Janus}, the existence of Janus solutions requires non-vanishing pseudoscalars resulting in more complicated set of BPS equations in constrast to the simple flat domain wall or RG flow case. We begin with the conditions $\delta \psi^i_{\hat{\mu}}=0$ for $\hat{\mu}=0,1$ which give
\begin{equation}
A'\gamma_{\hat{r}}\epsilon_A+\frac{1}{\ell}e^{-A}\gamma_{\hat{\xi}}\epsilon_{A}-\mc{W}\epsilon^A=0\, .\label{dPsi_mu_eq}
\end{equation}
This leads to the following BPS equation
\begin{equation}
A'^2=W^2-\frac{1}{\ell^2}e^{-2A}\label{dPsi_BPS_eq}
\end{equation}
for $W=|\mc{W}|$. We still use the $\gamma_{\hat{r}}$ projection given in \eqref{gamma_r_Pro}. Imposing the $\gamma_{\hat{\xi}}$ projection of the form
\begin{equation}
\gamma_{\hat{\xi}}\epsilon_A=i\kappa e^{i\Lambda}\epsilon^A\label{gamma_xi_pro}
\end{equation}
with $\kappa^2=1$, we can solve the condition \eqref{dPsi_mu_eq} for the $\gamma_{\hat{r}}$ projector leading to the phase factor
\begin{equation}
e^{i\Lambda}=\frac{A'}{W}+\frac{i\kappa}{\ell}\frac{e^{-A}}{W}\label{real_W_phase}
\end{equation}
for real $\mc{W}$ and
\begin{equation}
e^{i\Lambda}=\frac{\mc{W}}{A'+\frac{i\kappa}{\ell}e^{-A}}\label{complex_W_phase}
\end{equation}
for complex $\mc{W}$. We also note that the constant $\kappa=\pm 1$ corresponds to the chiralities of the Killing spinors on the two-dimensional conformal defects described by the $AdS_3$-slices. Finally, the conditions $\delta \psi^A_{\hat{\xi}}=0$ and $\delta\psi^A_{\hat{r}}=0$ can be solved to obtain the explicit form of the Killing spinors, see details in \cite{warner_Janus},
\begin{equation}
\epsilon_A=e^{\frac{A}{2}+\frac{\xi}{2\ell}+i\frac{\Lambda}{2}}\varepsilon^{(0)}_A
\end{equation}
in which the constant spinors $\varepsilon^{(0)}_{A}$ could possibly have an $r$-dependent phase and satisfy
\begin{equation}
\gamma_{\hat{r}}\varepsilon^{(0)}_{A}=\varepsilon^{(0)A}\qquad
\textrm{and}\qquad
\gamma_{\hat{\xi}}\varepsilon^{(0)}_{A}=i\kappa\varepsilon^{(0)A}\, .
\end{equation}
\indent It turns out that among the previously considered cases only $SO(2)\times SO(4)$ and $SO(3)$ symmetric scalars can possibly possess supersymmetric Janus solutions. This is mainly a consequence of the consistency in turning on non-vanishing pseudoscalars. For the $SO(3)$ case, the analysis is highly complicated as already seen in the case of RG flows considered in the previous section. Therefore, we will only give the Janus solution with $SO(2)\times SO(4)$ symmetry. This case is more traceable, and it turns out that the solution can be analytically obtained.

\subsection{Janus solutions with $SO(2)\times SO(4)$ symmetry}
In this case, the superpotential is real, so we will use the phase $e^{i\Lambda}$ from \eqref{real_W_phase}. In general, since $\epsilon^{1,2}$ and $\epsilon^A$ with  $A=3,4,5,6$ transform differently under $SO(2)\times SO(4)$ namely as $(\mathbf{2},\mathbf{1})+(\mathbf{1},\mathbf{4})$, the two sets of Killing spinors can satisfy different projectors. We find that in order to obtain a consistent set of BPS equations we need to choose opposite signs of $\kappa$ for $\epsilon^{1,2}$ and $\epsilon^{3,\ldots, 6}$. Therefore, the surface defect will preserve $N=(2,4)$ or $N=(4,2)$ superconformal symmetry.
\\
\indent With the superpotential
\begin{equation}
\mc{W}=4g\cosh \varphi,
\end{equation}
we find the following BPS equations
\begin{eqnarray}
\varphi'&=&-\frac{8g^2\ell^2 A'e^{2A}\sinh(2\varphi)}{1+\ell^2A'^2e^{2A}},\label{JanusSO4_varphi_eq}\\
\zeta'&=&-\frac{16g^2\kappa\ell e^{A}}{1+\ell^2A'^2e^{2A}},\label{JanusSO4_zeta_eq}\\
A'^2+\frac{e^{-2A}}{\ell^2}&=&16g^2\ell^2\cosh^2\varphi\, .\label{JanusSO4_A_eq}
\end{eqnarray}
It should be noted that, for $\ell\rightarrow \infty$, these equations reduce to those of the RG flow studied in the previous section. Furthermore, these equations take a very similar form to the $SO(4)$ symmetric Janus solution in $N=5$ gauged supergravity studied in \cite{N5_flow}.
\\
\indent By taking $\varphi$ as an independent variable, we can solve for $A(\varphi)$ and $\zeta(\varphi)$. The complete solution is given by
\begin{eqnarray}
A&=&C-\ln \sinh\varphi,\\
\cosh(2\varphi)&=&\frac{32g^2\ell^2\tanh^2[4g(r-r_0)]}{16g^2\ell^2-1},\\
\kappa \tan\zeta&=&-\sqrt{1-16g^2\ell^2}\sinh[4g(r-r_0)]
\end{eqnarray}
for constants $C$ and $r_0$. This solution takes the same form as the solution given in \cite{warner_Janus}, \cite{N5_flow} and \cite{N3_Janus} in $N=8,5,3$ gauged supergravities. We also note the unbroken supersymmetries on the conformal defects in these cases as follows: $N=(4,4)$, $N=(4,1)$ and $N=(2,1)$. All of these solutions should be related by truncations of $N=8$ gauged supergravity to $N=3$ and $N=5,6$ theories. This indicates that the $N=(4,4)$ Janus solution of \cite{warner_Janus} survives in the truncation to $N=3,5,6$ gauged supergravities. According to the AdS/CFT correspondence, we then expect the dual $N=3,5,6$ SCFTs to possess the same two-dimensional conformal defect as in the $N=8$ SCFT.
\\
\indent We end this section by giving a brief comment on the possible $SO(3)$ symmetric Janus solution. A partial analysis shows that there appears to be no obstruction in obtaining the BPS equations for Janus solutions in this case as in a similar analysis of $N=5$ theory in \cite{N5_flow}. Therefore, we expect a supersymmetric $N=2$ Janus solution with $SO(3)$ symmetry to exist in $N=6$ gauged supergravity as well. Since the analysis is far more complicated than the $SO(2)\times SO(4)$ case, we refrain from giving any definite result here.

\section{Supersymmetric $AdS_4$ black holes}\label{AdS4_BH}
In this section, we look for supersymmetric $AdS_4$ black holes in the form of curved-domain wall solutions interpolating between a locally asymptotically $AdS_4$ and $AdS_2\times \Sigma^2$ geometries. The $AdS_2\times \Sigma^2$ space describes the near horizon geometry of the black holes. In the following analysis, we will consider only the cases of $\Sigma^2$ being a two-sphere $(S^2)$ and a hyperbolic space $(H^2)$.
\\
\indent The metric ansatz is taken to be
\begin{equation}
    ds^2 = - e^{2 f(r)}dt^2 + dr^2 + e^{ 2 h(r)} (d\theta^2+F^2(\theta)d\phi^2)
\end{equation}
with the function $F(\theta)$ defined by
\begin{equation}
    F(\theta) =
    \begin{cases}
    \sin \theta, &\,\Sigma_2 = S^2 \\
    \sinh \theta, & \,\Sigma_2 = H^2
    \end{cases}\, .
\end{equation}
It is straightforward to derive non-vanishing components of the spin connection
\begin{eqnarray}
    \omega^{\hat{t}\hat{r}} &=& f' e^{\hat{t}},\qquad    \omega^{\hat{\theta}\hat{r}} = h' e^{\hat{\theta}}, \nonumber \\
    \omega^{\hat{\phi}\hat{r}} &=& h' e^{\hat{\phi}},\qquad
    \omega^{\hat{\theta}\hat{\phi}} = \frac{F'}{F} e^{-h} e^{\hat{\phi}}
\end{eqnarray}
with $F'(\theta)=\frac{dF}{d\theta}$.
\\
\indent In general, the curvature of the $\Sigma^2$ part on the
world-volume of the domain wall will completely break supersymmetry. However, it is well-known that some amount of
supersymmetry can be preserved by performing a topological twist.
This can be achieved by turning on some gauge fields along
$\Sigma^2$ in such a way that the corresponding spin connection,
$\omega^{\hat{\theta}\hat{\phi}}$ in the above metric, is cancelled.
This turns the covariant derivative of $\epsilon_A$ along $\Sigma^2$ into a partial
one. The resulting Killing spinors are accordingly given by spinors
that are independent of the $\Sigma^2$ coordinates. 

\subsection{$SO(2)\times SO(2)\times SO(2)$ symmetric solutions}
We first consider the $SO(2)\times SO(2)\times SO(2)$ twist with the
following ansatz for the gauge fields
\begin{eqnarray}
& &A^{12}=A_1(r)dt-p_1F'(\theta)d\phi,\nonumber \\
& & A^{34}=A_2(r)dt-p_2F'(\theta)d\phi,\nonumber \\
& &A^{56}=A_3(r)dt-p_3F'(\theta)d\phi\, .
\end{eqnarray}
The constants $p_{i}$, $i=1,2,3$, are identified with magnetic
charges. The corresponding field strengths are given by
\begin{eqnarray}
& &F^{12}=A_1'dr\wedge dt+\kappa p_1F(\theta)d\theta\wedge d\phi,\nonumber \\ 
& &F^{34}=A_2'dr\wedge dt+\kappa p_2F(\theta)d\theta\wedge d\phi,\nonumber \\ 
& &F^{56}=A_3'dr\wedge dt+\kappa p_3F(\theta)d\theta\wedge d\phi\, .
\end{eqnarray}
We have introduced a parameter $\kappa$ with $\kappa=1,-1$ for
$\Sigma^2$ being $S^2$ or $H^2$, respectively. We also note that
$F''(\theta)=-\kappa F(\theta)$.
\\
\indent With the $SO(2)\times SO(2)\times SO(2)$ singlet scalars
given by \eqref{SO2_3_singlet}, we find non-vanishing components of the
composite connection
\begin{equation}
{Q_{A}}^B=2gi\sigma_2\otimes\left(\begin{array}{ccc}
A^{12} & & \\
 & A^{34} &  \\
 & & A^{56} \\
\end{array}\right).
\end{equation}
With the component ${Q_{\hat{\phi}A}}^B$, the spin connection $\omega^{\hat{\theta}\hat{\phi}}$ can be
cancelled by imposing the following projector
\begin{equation}
\gamma_{\hat{\theta}\hat{\phi}}\epsilon_A={(i\sigma_2\otimes
\mathbb{I}_3)_A}^B\epsilon_B\label{gamma_th_ph_proj}
\end{equation}
and the twist conditions
\begin{equation}
2gp_1=2gp_2=2gp_3=1\, .
\end{equation}
These conditions imply that $p_1=p_2=p_3$. The twist is then obtained from the diagonal subgroup $SO(2)_{\textrm{diag}}\subset SO(2)\times SO(2)\times SO(2)$ as in the pure $N=4$ and $N=5$ gauged supergravities studied in \cite{flow_acrossD_bobev} and \cite{N5_flow}, respectively. For consistency, we will also set $A_3=A_2=A_1$. 
\\
\indent For all $p_i$ non-vanishing, the twists allow all the supersymmetries corresponding to $\epsilon^A$,
$A=1,2,\ldots,6$ to be unbroken subject to the projector \eqref{gamma_th_ph_proj}. We also note some useful relations for deriving the full set of BPS
equations. Using $\epsilon_{\hat{0}\hat{r}\hat{\theta}\hat{\phi}}=1$
and $\gamma_5\epsilon_A=-\epsilon_A$, we find
\begin{equation}
\gamma^{\hat{0}\hat{r}}\epsilon_A=-i\gamma^{\hat{\theta}\hat{\phi}}\epsilon_A={(\sigma_2\otimes
\mathbb{I}_3)_A}^B\epsilon_B\, .\label{gamma_0r_proj}
\end{equation}
\indent It turns out that, we need to turn on the
$SO(2)\sim U(1)$ gauge field of $U(6)\sim SU(6)\times U(1)$ in order
to find a consistent set of BPS equations. We similarly take the
ansatz for this $U(1)$ gauge field to be
\begin{equation}
A^0=A_0(r)dt-p_0 F'(\theta)d\phi\qquad \textrm{and}\qquad F^0=A_0'(r)dr\wedge dt+\kappa
p_0d\theta\wedge d\phi\, .\label{A0_ansatz}
\end{equation}
We also note that both $A^0$ and $A^{IJ}$ can appear in the BPS
equations due to the off-diagonal element of the scalar coset
representative. In particular, we have the relations
\begin{eqnarray}
& &\hat{F}^+_{AB}=h_{\Lambda
AB}F^{+\Lambda}=h_{0AB}F^{+0}+\frac{1}{2}h_{IJ,AB}F^{+IJ},\\
& &\hat{F}^+=h_{\Lambda
0}F^{+\Lambda}=h_{00}F^{+0}+\frac{1}{2}h_{IJ0}F^{+IJ}\, .
\end{eqnarray}
However, $A^0$ does not participate in the twist since $\epsilon_A$ are not charged under the $U(1)$ factor outside $SU(6)$.
\\
\indent It is also useful to define the ``central charge'' matrix
\begin{equation}
\mc{Z}_{AB}=-\frac{1}{\sqrt{2}}(\hat{F}^+_{\hat{\theta}\hat{\phi}AC}-i\hat{F}^+_{\hat{0}\hat{r}AC}){(i\sigma_2\otimes
\mathbb{I}_3)^C}_B\, .
\end{equation}
In the present case, it turns out that $\mc{Z}_{AB}$ is proportional
to the identity matrix, $\mc{Z}_{AB}=\mc{Z}\delta_{AB}$.
\\
\indent With all these and the projector \eqref{gamma_r_Pro}, we
find the following BPS equation, from $\delta \psi_{\hat{\theta}A}$
and $\delta \psi_{\hat{\phi}A}$,
\begin{equation}
 h'e^{i\Lambda}-\mc{W}-\mc{Z}=0
\end{equation}
which leads to
\begin{equation}
h'=\pm |\mc{W}+\mc{Z}|\qquad \textrm{and}\qquad e^{i\Lambda}=\pm \frac{\mc{W}+\mc{Z}}{|\mc{W}+\mc{Z}|}\, .
\end{equation}
\indent With the projectors \eqref{gamma_r_Pro} and \eqref{gamma_0r_proj}, the condition $\delta\psi_{\hat{0}A}=0$ gives
\begin{equation}
(f'+2igA_1)e^{i\Lambda}-\mc{W}+\mc{Z}=0
\end{equation}
which implies
\begin{equation}
f'=\textrm{Re}[e^{-i\Lambda}(\mc{W}-\mc{Z})]\qquad \textrm{and}\qquad 2gA_1=\textrm{Im}[e^{-i\Lambda}(\mc{W}-\mc{Z})]\, .
 \end{equation}
The latter fixes the time component of the gauge fields. Finally, as in the case of domain walls and Janus solutions, the
condition $\delta\psi_{\hat{r}A}=0$ determines the $r$-dependence of
the Killing spinors giving rise to $\epsilon_A=e^{\frac{f}{2}}\epsilon_{A(0)}$. 
\\
\indent Similar to the RG flow case, it turns out that we need to set
$\zeta_1=\zeta_2=\zeta_3=0$ for consistency. This gives real
$\mc{W}$ and $\mc{Z}$ resulting in $e^{i\Lambda}=\pm 1$ and $A_1=0$. We will also set $A_0(r)=0$ for simplicity although it is not constrained by the previously obtained conditions. In addition, the compatibility between the BPS equations coming from
$\delta \chi_A$ and $\delta\chi_{ABC}$ requires
\begin{equation}
p_0=\kappa p_1\, .
\end{equation}
We also note that, in this case, the choice $p_0=0$ breaks all
supersymmetry. This implies that the $SO(2)\times SO(2)\times SO(2)$
twist needs to be accompanied by the $U(1)$ gauge field $A^0_\mu$.
\\
\indent With all these, we find a consistent set of BPS equations
given by
\begin{eqnarray}
\varphi_1'&=&-\frac{\pd |\mc{W}+\mc{Z}|}{\pd \varphi_1}\nonumber\\
&=&\frac{1}{2}e^{-\varphi_1-\varphi_2-\varphi_3}\left[2g(1+e^{2(\varphi_2+\varphi_3)}
-e^{2(\varphi_1+\varphi_2)}-e^{2(\varphi_1+\varphi_3)})\right. \nonumber \\
& &\left.-p_1\kappa e^{-2h+2\varphi_1+2\varphi_2+\varphi_3}\right],\\
\varphi_2'&=&-\frac{\pd |\mc{W}+\mc{Z}|}{\pd \varphi_2}\nonumber\\
&=&\frac{1}{2}e^{-\varphi_1-\varphi_2-\varphi_3}\left[2g(1-e^{2(\varphi_2+\varphi_3)}
-e^{2(\varphi_1+\varphi_2)}+e^{2(\varphi_1+\varphi_3)})\right. \nonumber \\
& &\left.-p_1\kappa
e^{-2h+2\varphi_1+2\varphi_2+\varphi_3}\right],\\
\varphi_3'&=&-\frac{\pd |\mc{W}+\mc{Z}|}{\pd \varphi_3}\nonumber\\
&=&\frac{1}{2}e^{-\varphi_1-\varphi_2-\varphi_3}\left[2g(1-e^{2(\varphi_2+\varphi_3)}
+e^{2(\varphi_1+\varphi_2)}-e^{2(\varphi_1+\varphi_3)})\right. \nonumber \\
& &\left.-p_1\kappa
e^{-2h+2\varphi_1+2\varphi_2+\varphi_3}\right],\\
h'&=&|\mc{W}+\mc{Z}|\nonumber \\
&=&\frac{1}{2}e^{-\varphi_1-\varphi_2-\varphi_3}\left[2g(1+e^{2(\varphi_2+\varphi_3)}
+e^{2(\varphi_1+\varphi_2)}+e^{2(\varphi_1+\varphi_3)})\right. \nonumber \\
& &\left.+p_1\kappa
e^{-2h+2\varphi_1+2\varphi_2+\varphi_3}\right],\\
f'&=&|\mc{W}-\mc{Z}|\nonumber \\
&=&\frac{1}{2}e^{-\varphi_1-\varphi_2-\varphi_3}\left[2g(1+e^{2(\varphi_2+\varphi_3)}
+e^{2(\varphi_1+\varphi_2)}+e^{2(\varphi_1+\varphi_3)})\right. \nonumber \\
& &\left.-p_1\kappa
e^{-2h+2\varphi_1+2\varphi_2+\varphi_3}\right].
\end{eqnarray}
For an $AdS_2\times \Sigma^2$ fixed point to exist, we require that
$\varphi_1'=\varphi_2'=\varphi_3'=h'=0$ and
$f'\sim\frac{1}{L_{AdS_2}}$. It can be easily verified that the
above equations do not admit any $AdS_2\times \Sigma^2$ fixed
points.
\\
\indent Although there is no supersymmetric $AdS_2\times \Sigma^2$ fixed point, we are able to analytically obtain the complete solution to these BPS equations. Since it might be useful for some holographic studies, we will present the solution here. By changing to a new radial coordinate $\rho$ using the relation $\frac{d\rho}{dr}=e^{\varphi_3}$, we can form the following linear combinations 
\begin{eqnarray}
\frac{d}{d\rho}(\varphi_1-\varphi_2)&=&2g(e^{\varphi_2-\varphi_1}-e^{\varphi_1-\varphi_2})\\
\textrm{and}\qquad \frac{d}{d\rho}(\varphi_2-\varphi_3)&=&2g(e^{\varphi_1-\varphi_2}-e^{\varphi_1+\varphi_2-2\varphi_3}).
\end{eqnarray}
 The first equation can be solved by
 \begin{equation}
 \varphi_1=\ln\left[\frac{e^{\varphi_2}(e^{4g\rho}+e^{4g\rho_0})}{e^{4g\rho}-e^{4g\rho_0}}\right]
 \end{equation}
 with an integraton constant $\rho_0$. Using this result in the second equation, we find the solution
 \begin{equation}
 \varphi_2=\ln \left[\frac{e^{\varphi_3-2g\rho}(e^{4g\rho}-e^{4g\rho_0})}{\sqrt{e^{4g\rho}+e^{8g\rho_0}+8gC}}\right]
 \end{equation}
with another integration constant $C$.
\\
\indent By treating $f$ and $h$ as functions of $\varphi_3$, we find the following solutions for $f$ and $h$
\begin{eqnarray}
f&=&-\frac{1}{2}\ln\left[e^{4g\rho_0}\left(256g^3C\tilde{C}-16g\tilde{C}e^{8g\rho_0}+\kappa p_1\ln\left[\frac{1+e^{8g(\rho-\rho_0)}+8gCe^{4g\rho-8g\rho_0}}{e^{8g(\rho-\rho_0)}-1}\right]\right)\right. \nonumber \\
& &\left.-8gC\kappa p_1\tanh^{-1}e^{4g(\rho-\rho_0)}\right]+h,\\
h&=&\frac{1}{2}\ln\left[e^{4g\rho_0}\left(16g\tilde{C}(e^{8g\rho_0}-16g^2C^2)-\kappa p_1\ln\left[\frac{1+e^{8g(\rho-\rho_0)}+8gCe^{4g\rho-8g\rho_0}}{e^{8g(\rho-\rho_0)}-1}\right]\right)\right.\nonumber \\
& &\left.\phantom{\frac{1}{e^8}}+8gC\kappa p_1\tanh^{-1}e^{4g(\rho-\rho_0)} \right]+\frac{1}{2}\ln\left[\frac{e^{12g\rho_0}(1-e^{8g(\rho-\rho_0)})^2}{8g(e^{8g\rho_0}-16g^2C^2)}\right]+\varphi_3-4g\rho\, .\qquad\quad
\end{eqnarray}
\indent Finally, the solution $\varphi_3(\rho)$ can be given implicitly in the following equation
\begin{eqnarray}
4C_0e^{4g\rho}(e^{8g\rho_0}+e^{8g\rho}+8gCe^{4g\rho})&=&\beta_0+\beta_1\ln \left[\frac{e^{4g(\rho_0-\rho)}+1}{e^{4g(\rho_0-\rho)}-1}\right]\nonumber \\
& &+\beta_2\ln\left[\frac{e^{8g(\rho-\rho_0)}-1}{1+e^{8g(\rho-\rho_0)}+8gCe^{4g\rho}}\right] 
\end{eqnarray}
in which $C_0$ is a constant, and the coefficients $\beta_0$, $\beta_1$ and $\beta_2$ are defined by
\begin{eqnarray}
\beta_0&=&-16g\tilde{C}e^{4\rho_0}(16Cg^2-e^{8g\rho_0})\left[2e^{4\varphi_3+8g(\rho+\rho_0)}+8gCe^{4g\rho}(e^{8g\rho}+e^{8g\rho_0})\right.\nonumber \\
& &\left.+(e^{16g\rho_0}+e^{16g\rho})(1-e^{4\varphi_3})+2e^{8g(\rho+\rho_0)} \right],\\
\beta_1&=&\left[\frac{\kappa p_1}{2(e^{4g\rho}+e^{4g(3\rho-2\rho_0)}+8Cge^{8g(\rho-\rho_0)})}\right]\left[e^{12g\rho}+e^{4g(\rho+2\rho_0)}+4Cge^{8g\rho}(3+e^{4\varphi_3})\right. \nonumber \\
& &\left.+16g^2C^2 e^{4g\rho}(1+e^{8g(\rho-\rho_0)})-2Cg(e^{4\varphi_3}-1)(e^{8g\rho_0}+e^{8g(2\rho-\rho_0)}) \right],\\
\beta_2&=&\frac{\kappa p_1[(e^{8g\rho_0}+e^{8g\rho}+8Cge^{4g\rho})^2-e^{4\varphi_3}(e^{8g\rho_0}-e^{8g\rho})^2]}{4e^{4g(\rho-\rho_0)}(e^{8g\rho_0}+e^{8g\rho}+8Cge^{4g\rho})}\, .
\end{eqnarray}
 \indent Since there is no $AdS_2\times \Sigma^2$ fixed point in the IR, the solution describes a flow from the locally supersymmetric $AdS_4$ vacuum to a curved domain wall with world-volume $\mathbb{R}\times \Sigma^2$. According to the AdS/CFT correspondence, the solution is expected to describe an RG flow from the $N=6$ SCFT in three dimensions to a supersymmetric quantum mechanics in the IR. The latter arises from the former by a twisted compactification on $\Sigma^2$. 
 
\subsection{$SO(2)\times SO(4)$ symmetric solutions}
We now look at a truncation of the previous result by setting
$p_2=p_3=0$ and $\varphi_2=\varphi_3=0$. The resulting solutions will preserve $SO(2)\times SO(4)$ symmetry with the twist performed along the $SO(2)$ factor. In this case, the
supersymmetry corresponding to $\epsilon^{3,4,5,6}$ will be broken
since we cannot perform a twist along these directions. With
$\epsilon^{3,4,5,6}=0$, we find the BPS equations
\begin{eqnarray}
\varphi'&=&\frac{1}{4}e^{-2h-\varphi}[8ge^{2h}-p_0+\kappa
p_1-e^{2\varphi}(8ge^{2h}+p_0+\kappa p_1)],
\\
h'&=&\frac{1}{4}e^{-2h-\varphi}[8ge^{2h}-p_0+\kappa
p_1+e^{2\varphi}(8ge^{2h}+p_0+\kappa p_1)],
\\
f'&=&\frac{1}{4}e^{-2h-\varphi}[8ge^{2h}+p_0-\kappa
p_1+e^{2\varphi}(8ge^{2h}-p_0-\kappa p_1)]
\end{eqnarray}
in which we have set $\varphi_1=\varphi$. We also note that with
only the $SO(2)$ twist, it is not necessary to set $p_0=\kappa p_1$.
However, the existence of an $AdS_2\times \Sigma^2$ fixed point
requires vanishing $p_0$. For $p_0=0$, we find a fixed point
\begin{equation}
\varphi=\varphi_0,\qquad h=\frac{1}{2}\ln\left[-\frac{\kappa
p_1}{8g}\right],\qquad L_{AdS_2}=\frac{1}{8g\cosh2\varphi_0}\, .
\end{equation}
for constant $\varphi_0$. This is an $AdS_2\times H^2$ fixed point
since the reality of $h$ implies $\kappa=-1$. 
\\
\indent The complete flow solution can be obtained by using the same
procedure as in the previous sections. The resulting solution is
given by
\begin{eqnarray}
h&=&\varphi-\ln(1-e^{2\varphi})+C,\\
f&=&h-2\varphi+\ln[\kappa
p_1(1+e^{4\varphi})+2e^{2\varphi}(4g-\kappa p_1)],\\
8g(\rho-\rho_0)&=&2\sqrt{\frac{2g}{\kappa
p_1-2g}}\tan^{-1}\left[\frac{4g+\kappa
p_1(e^{2\varphi}-1)}{2\sqrt{2g(\kappa
p_1-2g)}}\right]\nonumber \\
& &+\ln\left[\frac{\kappa
p_1(1+e^{4\varphi})+2e^{2\varphi}(4g-\kappa
p_1)}{(1-e^{2\varphi})^2}\right]
\end{eqnarray}
in which we have defined the new radial coordinate $\rho$ by
$\frac{d\rho}{dr}=e^{\varphi}$. We have neglected the integration
constant of $f$ by absorbing it in the rescaling of the time
coordinate $t$.
\\
\indent Near $r\sim \rho\rightarrow \infty$, we find
\begin{equation}
\varphi\sim e^{-4gr},\qquad h\sim f\sim 4gr
\end{equation}
which gives an asymptotically locally $AdS_4$ critical point. On the other hand, by choosing
$\varphi_0=\frac{1}{2}\ln\left(1-2\sqrt{-\frac{2g}{\kappa
p_1}}\right)$ and $C=-\varphi_0$, we find that as
$\varphi\rightarrow \varphi_0$
\begin{equation}
h\sim \frac{1}{2}\ln\left[-\frac{\kappa p_1}{8g}\right]\qquad
\textrm{and}\qquad f\sim 8gr\frac{1-\sqrt{-\frac{2g}{\kappa
p_1}}}{\sqrt{1-2\sqrt{-\frac{2g}{\kappa p_1}}}}
\end{equation}
which is the $AdS_2\times H^2$ fixed point identified above.
\\
\indent It should be noted that, in this case, the solution can be regarded as a solution of a truncated $N=2$ gauged supergravity. In particular, the solution with vanishing scalar corresponds to a universal RG flow across dimension of which the uplifts to M-theory and massive type IIA theory have been extensively studied in \cite{Zaffaroni_BH_entropy}. 

\subsection{$U(3)$ symmetric solutions}
As a final case, we consider $U(3)$ symmetric solutions with a twist performed along the $SO(2)\sim U(1)$ factor. The corresponding gauge generator of this $U(1)$ factor is given by $X_{14}+X_{25}+X_{36}$. We then turn on the following gauge fields 
\begin{equation}
\mc{A}=A^{14}=A^{25}=A^{36}=A(r)dt-\kappa p F'(\theta)d\phi\, .
\end{equation}  
With the $U(3)$ singlet scalar given in \eqref{U3_singlet}, we find the composite connection
\begin{equation}
{Q_{A}}^B=2gi\mc{A}{(\mathbb{I}_3\otimes \sigma_2)_A}^B\, .
\end{equation}
The twist is implemented by imposing 
\begin{equation}
\gamma_{\hat{\theta}\hat{\phi}}\epsilon_A={(\mathbb{I}_3\otimes i\sigma_2)_A}^B\epsilon_B\qquad \textrm{and}\qquad 2gp=1\, .
\end{equation}
We also note that, similar to the $SO(2)\times SO(2)\times SO(2)$ twist, all $\epsilon_A$ can be non-vanishing. In addition, we also need non-vanishing $A^0$ which we will again use the ansatz \eqref{A0_ansatz}. In this case, consistency requires $p_0=-\kappa p$.
\\
\indent As in the RG flow case, we need to set $\zeta=0$ for consistency between the BPS equations and the field equations. This again results in $A(r)=0$. Repeating the same analysis as in the previous cases, we find the following BPS equations
\begin{eqnarray}
\varphi'&=&g(e^{-\varphi}-e^{3\varphi})+\frac{1}{2}\kappa pe^{-2h-3\varphi},\\
h'&=&g(3e^{-\varphi}+e^{3\varphi})+\frac{1}{2}\kappa pe^{-2h-3\varphi},\\
\varphi'&=&g(3e^{-\varphi}+e^{3\varphi})-\frac{1}{2}\kappa pe^{-2h-3\varphi}\, .
\end{eqnarray}
It is easily verified that there is no $AdS_2\times \Sigma^2$ fixed point in these equations. In this case, we are not able to obtain the analytic flow solution.

\section{Conclusions and discussions}\label{conclusion}
In this paper, we have studied $N=6$ gauged supergravity in four dimensions with $SO(6)$ gauge group. The gauged supergravity can be obtained from a truncation of the maximal $N=8$ theory with $SO(8)$ gauge group. There is a unique $N=6$ supersymmetric $AdS_4$ vacuum preserving the full $SO(6)$ gauge symmetry. This can be identified with $AdS_4\times CP^3$ geometry in type IIA theory dual to an $N=6$ SCFT in three dimensions. We have found a number of RG flow solutions with various symmetries from this $N=6$ SCFT to possible non-conformal phases in the IR. In particular, there is one solution, breaking the $SO(6)$ R-symmetry to $SO(2)\times SO(4)$, with unbroken $N=6$ Poincare supersymmetry. This is precisely in agreement with the field theory result on mass deformations of $N=6$ SCFTs given in \cite{N5_6_3D_SCFT}. Other solutions preserve $U(3)$, $SO(3)$ and $SO(2)\times SO(2)\times SO(2)$ symmetries. While most of the solutions preserve $N=6$ supersymmetry, in the case of $SO(3)$ symmetry, it is possible to find $N=2$ supersymmetric solutions. We have analytically given all of these solutions and also checked that, except for the $N=2$ solution, the resulting IR singularities are physical by the criterion given in \cite{Gubser_singularity}.  
\\
\indent We have also considered more complicated solutions by generalizing the flat domain walls to the curved ones. In the case of $AdS_3$-sliced domain walls, we have found a supersymmetric Janus solution, describing a two-dimensional conformal defect within the $N=6$ SCFT, with $SO(2)\times SO(4)$ symmetry and $N=(2,4)$ supersymmetry on the defect. The resulting solution takes the same form as those given in $N=8$, $N=5$ and $N=3$ gauged supergravities studied in \cite{warner_Janus}, \cite{N5_flow} and \cite{N3_Janus}. We therefore argue that these solutions are related to the $N=8$ solution by truncations. In order for Janus solutions to exist, it is necessary that pseudoscalars are non-vanishing as pointed out in \cite{warner_Janus}. It turns out that, among the remaining cases studied in this work, only the $SO(3)$ invariant sector could possibly admit supersymmetric Janus solutions.   
\\
\indent Futhermore, we have studied supersymmetric solutions of the form $AdS_2\times \Sigma^2$ and the interpolating solutions between these geometries and the $N=6$ $AdS_4$ vacuum. We have found one $AdS_2\times H^2$ fixed point with $SO(2)\times SO(4)$ symmetry from $SO(2)$ twist. The solution interpolating between this fixed point and the $AdS_4$ vacuum preserves two supercharges while the IR fixed point $AdS_2\times H^2$ has four supercharges. Holographically, this solution corresponds to an RG flow from the $N=6$ SCFT to superconformal quantum mechanics which is useful in computing black hole entropy along the ling of \cite{Zaffaroni_BH_entropy,BH_entropy_benini,BH_entropy_Passias}. 
\\
\indent For $SO(2)\times SO(2)\times SO(2)$ twist, the BPS equations are more complicated but admit no $AdS_2\times \Sigma^2$ fixed point. However, in this case, we are able to obtain a complete flow solution between the $AdS_4$ critical point to a curved domain wall with world-volume $\mathbb{R}\times \Sigma^2$ in the IR. The solution preserves $N=6$ supersymmetry in three dimensions, or twelve supercharges, and $SO(2)\times SO(2)\times SO(2)$ symmetry. This should be dual to a twisted compactification on $\Sigma^2$ of the UV $N=6$ SCFT to a supersymmetric quantum mechanics in the IR. We have also looked for $AdS_2\times \Sigma^2$ geometries from an $SO(2)\sim U(1)$ twist in the case of $U(3)$ symmetric solutions, but there do not exist any $AdS_2\times \Sigma^2$ fixed points.    
\\
\indent Since all the solutions presented here are fully analytic, we hope they could be useful in the study of gauge/gravity holography and other related aspects. We also note that most of the structures of the solutions are very similar to those of the $N=5$ gauged supergravity studied in \cite{N5_flow}. In particular, the $N=6$ Poincare supersymmetry in three dimensions is unbroken on the domain wall solutions if there are no non-vanishing pseudoscalars. Unlike in the $N=5$ theory, we are not able to find a definite conclusion on whether this is true in general due to a more complicated scalar coset manifold. However, many similarities in the structures of various types of supersymmetric solutions suggest that this should be the case.        
\\
\indent There are a number of directions to extend the present work which is clearly only the first-step in classifying supersymmetric solutions of $N=6$ gauged supergravity. First of all, it would be interesting to uplift the RG flow solutions to M-theory via the embedding in $N=8$ gauged supergravity which in turn can be obtained from a truncation of M-theory on $S^7$. The time component $g_{00}$ of the eleven-dimensional metric can be used to determine whether the four-dimensional singularities, in particular the $N=2$ case, are physically acceptable in M-theory using the criterion given in \cite{Maldacena_Nunez_nogo}. This would lead to a complete holograhic description of mass deformations of $N=6$ CSM theory and possible related M-brane configurations. 
\\
\indent We have only considered gauged supergravity with $SO(6)$ gauge group electrically embedded in the global $SO^*(12)$ symmetry. It would  be interesting to study magnetic and dyonic gaugings involving also magnetic gauge fields. In particular, performing a similar study in the case of $N=6$ gauged supergravity with the electric-magnetic phase $\omega$, see \cite{omega_N6} and \cite{omega_N8,omega_N8_2}, could be of particular interest since in the omega deformed $N=8$ theory, the structure of vacua and domain walls are much richer than the electric counterpart, see \cite{Guarino_BPS_DW,omega_vacua_1,omega_vacua_2,omega_DW_2} for more detail. In addition, the study of genuine $N=6$ gaugings which cannot be embedded in the $N=8$ theory is worth considering. In this case, the gaugings do not satisfy extra quadratic constraints coming from the truncation of the $N=8$ theory, see a discussion in \cite{twin_SUGRA}, so the corresponding solutions cannot be embedded in the maximal theory.

\begin{acknowledgments}
We would like to give a special thank to M. Trigiante for a valuable correspondence. We also thank C. Panwisawas for his help in finding some literature. This work is supported by The Thailand Research Fund (TRF) under grant RSA6280022.
\end{acknowledgments}

\end{document}